\documentclass[aps,prl,showpacs,twocolumn,groupedaddress,superscriptaddress,floatfix]{revtex4-1}

\usepackage[utf8]{inputenc} 
\usepackage[T1]{fontenc}
\usepackage{graphicx}
\usepackage{amssymb}
\usepackage{amsmath}
\usepackage{relsize}
\usepackage{color}

\begin{document}
\title{Optimal Placement of Origins for DNA Replication}

\author{Jens Karschau}
\affiliation{Institute for Complex Systems and Mathematical Biology, SUPA, King's College, University of Aberdeen, Aberdeen AB24 3UE, United Kingdom}
\affiliation{Wellcome Trust Centre for Gene Regulation and Expression, College of Life Sciences, University of Dundee, Dundee DD1 5EH, United Kingdom}
\email{jens.karschau@abdn.ac.uk}

\author{J. Julian Blow}
\affiliation{Wellcome Trust Centre for Gene Regulation and Expression, College of Life Sciences, University of Dundee, Dundee DD1 5EH, United Kingdom}

\author{Alessandro P. S. de Moura}
\affiliation{Institute for Complex Systems and Mathematical Biology, SUPA, King's College, University of Aberdeen, Aberdeen AB24 3UE, United Kingdom}

\date{\today}

\begin{abstract}
DNA replication is an essential process in biology and its timing must
be robust so that cells can divide properly. Random fluctuations in
the formation of replication starting points, called \emph{origins},
and the subsequent activation of proteins lead to
variations in the replication time. We analyse these stochastic
properties of DNA and derive the positions of origins corresponding to
the minimum replication time.  We show that under some conditions the
minimization of replication time leads to the grouping of origins,
and relate this to experimental data in a number of species showing
origin grouping.
\end{abstract}


\maketitle

\begin{figure}[tb]
 \begin{center}
\includegraphics[]{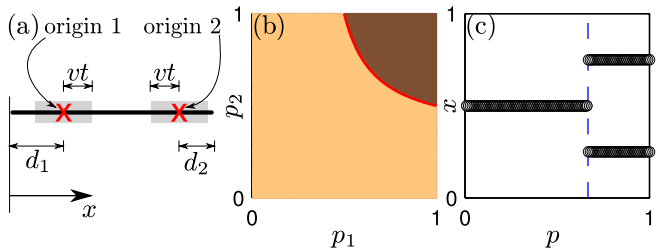} 	
\caption{\label{fig:Figure1} (color online).  (a)~Coordinate system for origin  loci with $d_1,\ d_2$ being the distance from the left- or right-end of the chromosome, respectively. $x$ is the position coordinate along the chromosome. Replication forks travel at a speed $v$ away from the origins. The grey regions show the replicated DNA at time $t$. (b)~Simulation results, showing optimal  loci to achieve minimal  $T_\text{rep}$ for 2 loci with different competences, are  shown for $p_1$-$p_2$ combinations on a lattice grid. Color indicates $d_1,d_2=1/2$ (beige) or $d_1,d_2=1/4$ (brown). The two regimes are separated  by a coexistence line  matched by the condition Eq.~(\ref{eq:2orivar}) in red. (c)~Optimal position of 2 loci  with respect to their competence $p$ to minimize the replication  time $T_\text{rep}$ (circles) and $p=2/3$  (dashed line).}
	\end{center}
\end{figure}

The replication of the DNA content of a cell is one of the most
important processes in living organisms. It ensures that the
information needed to synthesize proteins and cellular components is
passed on to daughter cells in a robust and timely fashion.
Replication takes place during the S-phase of the cell cycle, and
it starts from specific locations in the chromosome called
\emph{origins}. In order to function in a particular round of the cell
cycle, possible origin locations (loci) must undergo a sequence of binding events before
S-phase starts. This culminates in the clamping of one or more pairs of ring-shaped
{Mcm2-7} molecules  around the DNA; this is known as \emph{licensing}. Below we denote a pair of Mcm molecules as pMcm.
Features of human replication have been studied with the help of the yeast \emph{S. cerevisiae}  and \emph{X. laevis} frog embryos.  In
\emph{S. cerevisiae} licensing is only possible at a set of
specific points in each chromosome, characterized by the presence of
specific DNA sequences, whereas in \emph{X. laevis} embryos the licensing proteins can
bind at virtually any location in the genome~\cite{Kelly2000}. When a licensed locus activates
in S-phase, two replication forks are created at the origin, and
they move in opposite directions with approximately constant speed,
duplicating the DNA as they travel through the chromosome
[Fig.~\ref{fig:Figure1}(a)].  Both origin licensing and origin activation time are stochastic
events, since they result from molecular processes involving
low-abundance species.  In
\emph{X. laevis}, both the loci that are licensed and licensed loci selected for activation vary randomly from cell to cell,
whereas in \emph{S. cerevisiae} each of the fixed loci has
a certain probability of being activated in any given cell --- the
\emph{competence} --- which reflects the fraction of cells in a
population in which that locus has had time to be licensed
before the S-phase starts~\cite{Moura2010}.

The total time it takes to replicate a cell's DNA --- the
\emph{replication time} --- is a quantity of crucial importance for
biology, since it is clearly an evolutionary advantage for replication
to be rapid as it affects the minimum time required for cells to duplicate.  The location of the origins is one of the crucial
factors determining the replication time of cells, and it is
reasonable to expect that the loci have been selected by
evolution such that the replication time is minimized.  There
are a number of recent theoretical and modeling works on the dynamics
of DNA replication (reviewed in~\cite{Hyrien2010}).  Previous
theoretical works on \emph{S. cerevisiae} have used the experimentally
determined loci as given parameters, without attempting to
understand why the origins are located where they are~\cite{Spiesser2009,Bruemmer2010,Moura2010,Yang2010}.  Inspection of the loci on a \emph{S. cerevisiae} genome map shows
groups of two or three very close origins which are very prominent in most
chromosomes~\cite{Nieduszynski2007}. There is also experimental evidence for 
grouping in \emph{X. laevis}, where origins seem to be distributed
with groups of  5 to 10 pMcms~\cite{Mahbubani1997,Blow2001a,Edwards2002}.Most of the existing models of
replication in \emph{X. laevis}~\cite{Blow2001a,Jun2004a,Zhang2006,Goldar2008,Yang2008,Blow2009} -- an exception is~\cite{Jun2004b} -- assume the origins to be
random and independent of each other, and so they cannot explain 
pMcm grouping or the observed maximum-spacing of 25~kilobases (kb) between adjacent origins~\cite{Blow2001a}.

In this Letter, we use a simplified mathematical model of the DNA
replication process to determine the optimal origin location in a
chromosome which leads to the shortest average replication time, and how this
optimal placement depends on parameters such as the origin competences
and the width of the activation time probability distribution.  We
show that contrary to what one might expect, in many cases the
replication time is minimized by placing origins close together in
groups like those observed in real chromosomes.  This suggests that
origin locations have been selected to minimize the replication time.
Analysis of our model reveals that grouping is favored for
low-competence origins and for origins with large stochastic
fluctuations in their activation time.  The reason for this is that
if origins have an appreciable likelihood of either failing to
activate (low competence) or of taking a very long time to activate,
grouping origins together helps reduce the risk of large regions of
the chromosome  not being replicated on time.  If the
origins are highly competent and have a well-defined activation time,
it is optimal to have maximal coverage and distribute the origins evenly
on the chromosome.  We further show that there is an abrupt transition
in the optimal configuration of origins, from isolated to grouped,
as the locus' competence decreases (in \emph{S. cerevisiae}), and also as
the width of the activation time distribution increases.  We give an
intuitive explanation of this phenomenon, and argue that it is robust,
and independent of any particular details of the model.  These results
are derived analytically, and tested through numerical simulations.
We also compare quantitatively the predictions of our theory with the
available experimental data for both \emph{S. cerevisiae} and
\emph{X. laevis}, and find that they match well.

We start by analyzing the case of stochasticity in licensing of fixed origin loci, as in
\emph{S. cerevisiae}.  DNA is modeled as a one-dimensional segment of
unit length, and we for simplicity consider only two loci in the
chromosome.  The two origin loci have competences $p_1$ and $p_2$ ---
these are the probabilities that origins have been licensed and can
therefore start replication forks. We initially make the 
assumption that origins activate at a well-defined time (which we
set to $t=0$).  All replication forks travel at the same unit speed across the DNA. We consider the geometry depicted in
Fig.~\ref{fig:Figure1}(a);  $d_1\ (d_2)$ is the distance from the left (right) end of the chromosome to the left (right) most locus. If both loci fail to be licensed we postulate that replication
will eventually take place anyway, with a replication time $T_0$ ---
for example, we can imagine that stretch of DNA will be replicated by
forks originating from origins outside of the region we are considering.  Our results do not depend on $T_0$, as
will be clear shortly; this is just a mathematical device to prevent
us dealing with infinite replication times.

If only one of the loci fails to become licensed, the replication time
depends on the time it takes for the fork to reach the furthest end of
the segment, so $T_{d_1}=1-d_1$ for locus 1 and $T_{d_2}=1-d_2$ for
locus 2.  If both loci have been licensed the replication time
$T_{d_1,d_2}=\max\{d_1, d_2, (1-d_1-d_2)/2 \}$ is defined by the
longest time for a fork to reach the end of the segment or for two
forks to collide. It can be shown that the replication time of an asymmetric placement of
loci is never less than a corresponding symmetric configuration
(that is, with $d_1=d_2$).  Therefore we consider only
symmetrical locus placements, and use $d_1=d_2=d$ with $0\leq d\leq
1/2$. The average replication time is then given by $T_{\text{rep}}(d) = (1-p_1)(1-p_2)T_0 + (p_1+p_2-2p_1p_2)(1-d)+ p_1p_2\max\{d,(1-2d)/2\}$.

This is a piecewise-linear function with discontinuities at
$d=1/4$ and $1/2$.  Hence, $T_\text{rep}$ can only have a minimum
found at $d=0,\ d=1/2$, or at $1/4$. Placing loci at the end of a
segment ($d=0$) is obviously not a minimum of $T_\text{rep}$. The
replication times for $d=1/4$ and $1/2$ are $T_\text{rep}(d=1/2) = (1-p_1)(1-p_2)T_0 + \left(p_1
+ p_2 - p_1p_2\right)/2$ and $T_\text{rep}(d=1/4) = (1-p_1)(1-p_2)T_0 + \left(3p_1 + 3p_2 -
5p_1p_2\right)/4$.  We conclude that the two loci group together
($d=1/2$) to achieve minimum replication time if $T_\text{rep}(d=1/2)<T_\text{rep}(d=1/4)$,
which leads to the condition
\begin{equation}
p_2 < \frac{p_1}{3p_1-1}  \label{eq:2orivar}.
\end{equation}
Notice here that $T_0$ drops out.  The inequality
Eq.~(\ref{eq:2orivar}) defines two regions on the $p_1$--$p_2$ plane,
corresponding to grouped or isolated loci being optimum.  This is shown in Fig.~\ref{fig:Figure1}(b), where this
analytical result is confirmed by stochastic simulations. The region
above the curve corresponds to competences for which $T_{\text{rep}}$
is minimized by loci being apart ($d=1/4$) and below the curve
for organising these in a group ($d=1/2$). In general, if one of the loci
has low competence grouping gives the minimum $T_\text{rep}$.  In
fact, it can be shown that if one of the
loci has a competence lower than 50\%, grouping is the optimal
situation regardless of the competence of the other --- even if
the other is close to 100\% competent.

For the case of equal competences, $p_1=p_2=p$, the grouped
configuration is optimal if $p<2/3$. We ran a numerical optimization
algorithm (using genetic algorithms~\cite{PGAPack}) to find the loci
corresponding to the least $T_{\text{rep}}$ for a range of $p$; these results are shown in Fig. \ref{fig:Figure1}(c).  The same transition
also takes place for non-identical values of $p_1$ and $p_2$ ---
whenever one crosses from the dark to the light regions of
Fig. \ref{fig:Figure1}(b).

The above results may seem at first quite counter-intuitive; one might expect that the configuration with the least replication
time would correspond to isolated loci ($d=1/4$).  However, if the origins have a significant chance of
failing to activate, this configuration would mean that often one side
of the chromosome would have to wait for a fork which originated at the origin
on the other site to replicate it, therefore
increasing $T_{\text{rep}}$.  So in the case of low competences, it
becomes advantageous to have both loci centered, which is near any point in the chromosome.  This explains the condition for grouping if  $p<2/3$.

\begin{figure}[tb]
	{\includegraphics[]{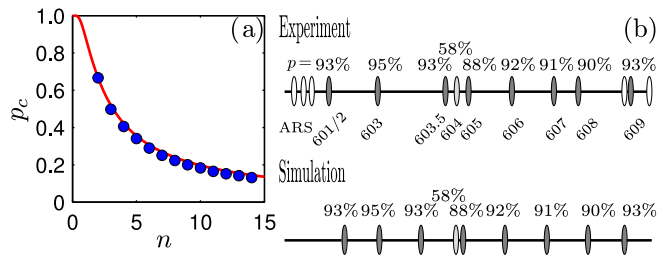}}
	\caption{(color online). (a) Probability at which groups separate $p_c$ vs. loci/group $n$. Shown are simulations (circles) and analytical prediction for $p_c = 1 - 1/ \sqrt[n]{9}$ (line). (b) Distribution of origin loci on yeast chromosome VI with known (grey) and unknown competences~\cite{Shirahige1993,Moura2010}. The distribution results from our simulation in search for minimum $T_\text{rep}$ (only grey origins considered). The group in the middle of the chromosome with a low and highly competent locus was recovered.}
	\label{fig:Figure2}
\end{figure}

In reality eukaryotic chromosomes have more than two loci
\cite{Hyrien2010}, so next we investigate the case of a chromosome
on which there are many loci and examine the conditions under which
it becomes favorable to have isolated origin loci compared to
groups.  In this analysis we will assume for simplicity that the
loci all have identical competence.  We consider a group of loci as one single locus with an effective competence
$p_\text{eff}$. For a group consisting of $m$ loci
$p_\text{eff}$ is the competence that at least one locus will be
licensed there, and is given by $p_{\text{eff}}=1-(1-p)^m$.
We assume that one large group of $n$ identical  loci breaks
up into two groups of equal size, each consisting  of $n/2$ loci. A locus organized with others in a group of size $m=n/2$ rather than with $n$ loci will give minimum $T_{\text{rep}}$, as long as the locus' competence is less than its
critical probability $p_c$, given by $p_{\text{eff}}=2/3$, which yields $
p_c = 1 - 1/ \sqrt[n]{9}$.
Figure~\ref{fig:Figure2}(a) confirms our analytical result showing the
value of $p_c$ for increasing group sizes in our simulations. These
results clearly show that large groups of many highly competent
loci are unfavorable, but that groups tend to form for
low-competence loci.

Our hypothesis is that selective pressure has influenced the position of 
origin loci through the minimization of the replication time.  We show in Fig.~\ref{fig:Figure2}(b) locus
competence and location data for yeast chromosome VI, which has been
studied
extensively~\cite{Shirahige1993,Moura2010}. Competences
cannot be measured for all loci (in white), because
either they are too close to the end of the chromosome or to an
adjacent locus. We performed a search for the optimal 
position for the loci in the region with known competences using a genetic
algorithm \cite{PGAPack}. We remark that there is an identifiability problem as all strong loci have $p\sim90\%$ and we therefore constrained the ordering during the  optimization. Although in
this result we do not consider inter-origin variations in the origin
activation time, the predicted locus distribution from these
simulations bears a good resemblance to the actual spacing with a score of $F=0.11$~\footnote{$F=\frac{1}{9} \sum_ {i=1}^{n=9}d_i^o/d_i^r$  is a measure of the difference between the gap distribution of the optimized and random cases. A gap is defined as the separation between the $i^\text{th}$ experimental locus position $p_i^e$ and that of the optimization  $p_i^o$: $d_i^o=|p_i^e-p_i^o|$. $d_i^r$ is akin; the average separation that arises from placing a locus uniformly randomly and $p_i^e$. $F=0$ means that the optimization fits the experimental loci positions perfectly; $F\sim1$ indicates no difference to that of a random placement.};
in particular we recover the group in the middle, in
which an origin locus with 58\% competence is placed next to one
with 88\% competence. Even multiple repeats of the optimisation algorithm produce minimum $T_\text{rep}$ solutions which have on average $F=0.12$.

\begin{figure}[tb]
	\includegraphics[]{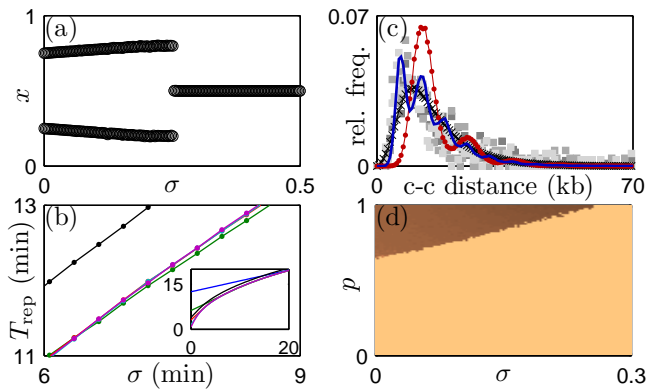}
	\caption{(color online). (a)~Origin position $x$ so that $T_\text{rep}$ for 2 pMcms is minimal on a segment of unit length, when the standard deviation $\sigma$ for their activation time increases.  (b) Inset: $T_\text{rep}$ as a function of $\sigma$ for realistic parameters as given in the text. Origins are distributed in 4 equally-spaced groups of 16 pMcms (blue);  8 groups of 8 pMcms (green); 16 groups of 4 pMcms (red); 32 groups of  2 pMcms (cyan);  64 single pMcms (magenta); 64 pMcms placed randomly (black). Main: zoom around realistic $\sigma\sim8$~min. For $6<\sigma<20$~min minimal $T_\text{rep}$  is achieved for groups of 8 pMcms.   (c) Center-center distribution from 3 experiments~\cite{Blow2001a} (squares)  and from simulations 5~min after replication started. The simulation is positioning groups of 4 pMcms every 6.3~kb (solid line), groups of 8 pMcms every 12.5~kb (circles), or all randomly (crosses). A small random amount was added to the group location of fixed distances which was picked from a Gaussian distribution with $\sigma\sim16\%$ of group distances. The $\text{pMcm}/\text{length}$ ratio was fixed  [cf.~(b)]. (d)~Phase diagram of the two-origin model to minimize replication time with changing competence and increasing the $\sigma$.  Color indicates origin position relative to chromosome ends $d_1,d_2=1/2$ (beige) or $d_1,d_2=1/4$ (brown).}
	\label{fig:Figure3}
\end{figure}

The above discussion focused on the case of pre-defined loci in yeast, and ignored additional noise such as the  variation in origin activation time. We show in Fig.~\ref{fig:Figure3}(d) that the previous pattern of origin grouping is preserved in the two-origin model with stochastic variation in origin activation time. Grouping is important for swift replication under conditions of low competence and large noise which we will explain in the remainder of this letter.

We now examine the case of stochastic activation time for  \emph{X. laevis} embryos as a model organism. Unlike loci in yeast, any DNA locus in an \emph{X. laevis} embryo is capable of binding with pMcm to become an origin.  Surprisingly, biologists find roughly equally-spaced groups of 5--10 pMcms separated by approximately 10~kb~\cite{Mahbubani1997,Blow2001a,Edwards2002}.  We will use the same approach as above, but now with respect to stochasticity in the replication
time.  In this case, an ``origin'' is defined as a locus where at least one pMcm
has bound to it, and so it corresponds to 100\% competent locus in the notation
we have used so far.  It is well accepted by biologists, however, that
origin activation time is stochastic.  For simplicity we assume
that the pMcms at an origin can activate with uniform probability at any time within a window which has a
lower boundary at $t_0=0$~min and an upper at $t_b$, which is
at maximum the length of an S-phase (20~min).  In addition,
pMcms are assumed to be all identical with the same activation
probability distribution (standard deviation $\sigma=t_b/\sqrt{12}$).

The expectation is that we will again see a transition of the optimal
configuration from isolated pMcms to groups as $\sigma$
increases; this is akin to varying competence in our previous scenario. We test this prediction using the
two-origin model with one pMcm bound to one origin;  we find numerically the optimal (minimum $T_\text{rep}$) positions for the origins as a function of $\sigma$.
The results are presented in Fig.~\ref{fig:Figure3}(a).  We again use
a segment of unit length and forks progress at unit speed of $v=1$.  We
observe a sharp transition at $\sigma\approx0.25$, above which it is best to
place both origins in the middle of the segment, as observed in
the case with varying competence.  A minor
difference between this case and the previous one is that for $\sigma<0.25$,
the optimal location of the origins is not constant.

We now apply this model for more origins and pMcms, using realistic parameters
so that we can relate the results to what is experimentally known
about \emph{X. laevis'} pMcm distribution.  We model a stretch of DNA
of size 100~kb and $v=1$ kb/min~\cite{Raghuraman2001}.  To determine
whether the minimum-replication-time configuration requires pMcm grouping, we distributed 64 pMcms in total, i.e. that there is on average $1/1.5~\text{pMcm}/\text{kb}$ as found in nature~\cite{Mahbubani1997}. The pMcms are then placed in
$64/{n}$ groups of $n\in\{1, 2, 4, 8, 16\}$ origins, so that the
origins are uniformly distributed through the 100~kb chromosome, or completely random. Other authors have identified $\sigma$ to be 6--10~min in
 \emph{X. laevis}~\cite{Herrick2002a, Goldar2008} as well as in \emph{S. cerevisiae} ~\cite{Raghuraman2001,Bruemmer2010,Moura2010,Sekedat2010}. Our results [Fig.~\ref{fig:Figure3}(b)] indicate that grouping with an equal spacing of up to 12.5~kb achieves
precise and fast DNA synthesis before the end of S-phase (20~min) for $\sigma$ within these limits. We also find that 8 groups of 8 pMcms gives the advantage of a 1.1~min quicker $T_\text{rep}$ than using random loci; even when the number of pMcms at these 8 groups varies, a quicker $T_\text{rep}$ is achieved (data not shown). Grouping pMcms also protects the overall replication process against fluctuations from one round of the cell cycle to another; a similar problem is discussed in~\cite{Yang2008}. This is because one initiation event at an origin is sufficient to activate replication forks.

One might expect that in a natural environment there would not be strict equal spacing of groups. We now relax our previous assumption by taking evenly-spaced groups and perturb the location of each group by a small random amount drawn from a Gaussian distribution.  The introduction of such variation allows us to compare our simulation with available experimental data of replicated genomic regions, which were captured as center-center distances at around 5~min after the onset of replication (cf. \cite{Blow2001a}). Figure~\ref{fig:Figure3}(c)  shows that our result is in agreement with the current understanding of the biological community, i.e. groups of 5-10~pMcms about every 10~kb. This may be achieved by a regulation of pMcm-loading proteins, whose affinity to bind decreases around existing origins \cite{Rowles1999, Oehlmann2004}. Although a random placement represents the data similarly well, $T_\text{rep}$ remains smaller in this case where the origin groups are not equally-spaced as seen before [cf. Figure~3(b)]. This shows that grouping of origins remains favorable even in a more general setting [Fig.~\ref{fig:Figure3}(d)].

\begin{acknowledgments}
We thank C. A. Nieduszynski and C. A. Brackley for critical reading of the manuscript. We also would like to thank the referees for their suggestions to the manuscript. This work has been
supported through the Scottish University Life Sciences Alliance and the Biotechnology and Biological Sciences Research
Council (grant numbers BB/G001596/1 and BB-G010722).
\end{acknowledgments}


\begin{thebibliography}{24}%
\makeatletter
\providecommand \@ifxundefined [1]{%
 \@ifx{#1\undefined}
}%
\providecommand \@ifnum [1]{%
 \ifnum #1\expandafter \@firstoftwo
 \else \expandafter \@secondoftwo
 \fi
}%
\providecommand \@ifx [1]{%
 \ifx #1\expandafter \@firstoftwo
 \else \expandafter \@secondoftwo
 \fi
}%
\providecommand \natexlab [1]{#1}%
\providecommand \enquote  [1]{``#1''}%
\providecommand \bibnamefont  [1]{#1}%
\providecommand \bibfnamefont [1]{#1}%
\providecommand \citenamefont [1]{#1}%
\providecommand \href@noop [0]{\@secondoftwo}%
\providecommand \href [0]{\begingroup \@sanitize@url \@href}%
\providecommand \@href[1]{\@@startlink{#1}\@@href}%
\providecommand \@@href[1]{\endgroup#1\@@endlink}%
\providecommand \@sanitize@url [0]{\catcode `\\12\catcode `\$12\catcode
  `\&12\catcode `\#12\catcode `\^12\catcode `\_12\catcode `\%12\relax}%
\providecommand \@@startlink[1]{}%
\providecommand \@@endlink[0]{}%
\providecommand \url  [0]{\begingroup\@sanitize@url \@url }%
\providecommand \@url [1]{\endgroup\@href {#1}{\urlprefix }}%
\providecommand \urlprefix  [0]{URL }%
\providecommand \Eprint [0]{\href }%
\providecommand \doibase [0]{http://dx.doi.org/}%
\providecommand \selectlanguage [0]{\@gobble}%
\providecommand \bibinfo  [0]{\@secondoftwo}%
\providecommand \bibfield  [0]{\@secondoftwo}%
\providecommand \translation [1]{[#1]}%
\providecommand \BibitemOpen [0]{}%
\providecommand \bibitemStop [0]{}%
\providecommand \bibitemNoStop [0]{.\EOS\space}%
\providecommand \EOS [0]{\spacefactor3000\relax}%
\providecommand \BibitemShut  [1]{\csname bibitem#1\endcsname}%
\let\auto@bib@innerbib\@empty
\bibitem [{\citenamefont {Kelly}\ and\ \citenamefont
  {Brown}(2000)}]{Kelly2000}%
  \BibitemOpen
  \bibfield  {author} {\bibinfo {author} {\bibfnamefont {T.~J.}\ \bibnamefont
  {Kelly}}\ and\ \bibinfo {author} {\bibfnamefont {G.~W.}\ \bibnamefont
  {Brown}},\ }\href {http://www.ncbi.nlm.nih.gov/pubmed/10966477} {\bibfield
  {journal} {\bibinfo  {journal} {Annu. Rev. Biochem.}\ }\textbf {\bibinfo
  {volume} {69}},\ \bibinfo {pages} {829} (\bibinfo {year} {2000})}\BibitemShut
  {NoStop}%
\bibitem [{\citenamefont {de~Moura}\ \emph {et~al.}(2010)\citenamefont
  {de~Moura} \emph {et~al.}}]{Moura2010}%
  \BibitemOpen
  \bibfield  {author} {\bibinfo {author} {\bibfnamefont {A.~P.~S.}\
  \bibnamefont {de~Moura}} \emph {et~al.},\ }\href {\doibase
  10.1093/nar/gkq343} {\bibfield  {journal} {\bibinfo  {journal} {Nucl. Acids
  Res.}\ }\textbf {\bibinfo {volume} {38}},\ \bibinfo {pages} {5623} (\bibinfo
  {year} {2010})}\BibitemShut {NoStop}%
\bibitem [{\citenamefont {Hyrien}\ and\ \citenamefont
  {Goldar}(2010)}]{Hyrien2010}%
  \BibitemOpen
  \bibfield  {author} {\bibinfo {author} {\bibfnamefont {O.}~\bibnamefont
  {Hyrien}}\ and\ \bibinfo {author} {\bibfnamefont {A.}~\bibnamefont
  {Goldar}},\ }\href {http://www.ncbi.nlm.nih.gov/pubmed/19936948} {\bibfield
  {journal} {\bibinfo  {journal} {Chromosome Res.}\ }\textbf {\bibinfo {volume}
  {18}},\ \bibinfo {pages} {147} (\bibinfo {year} {2009})}\BibitemShut
  {NoStop}%
\bibitem [{\citenamefont {Spiesser}\ \emph {et~al.}(2009)\citenamefont
  {Spiesser}, \citenamefont {Klipp},\ and\ \citenamefont
  {Barberis}}]{Spiesser2009}%
  \BibitemOpen
  \bibfield  {author} {\bibinfo {author} {\bibfnamefont {T.~W.}\ \bibnamefont
  {Spiesser}}, \bibinfo {author} {\bibfnamefont {E.}~\bibnamefont {Klipp}}, \
  and\ \bibinfo {author} {\bibfnamefont {M.}~\bibnamefont {Barberis}},\ }\href
  {\doibase 10.1007/s00438-009-0443-9} {\bibfield  {journal} {\bibinfo
  {journal} {Mol. Genet. Genomics}\ }\textbf {\bibinfo {volume} {282}},\
  \bibinfo {pages} {25} (\bibinfo {year} {2009})}\BibitemShut {NoStop}%
\bibitem [{\citenamefont {Br\"{u}mmer}\ \emph {et~al.}(2010)\citenamefont
  {Br\"{u}mmer} \emph {et~al.}}]{Bruemmer2010}%
  \BibitemOpen
  \bibfield  {author} {\bibinfo {author} {\bibfnamefont {A.}~\bibnamefont
  {Br\"{u}mmer}} \emph {et~al.},\ }\href {\doibase
  10.1371/journal.pcbi.1000783} {\bibfield  {journal} {\bibinfo  {journal}
  {PLoS Comput. Biol.}\ }\textbf {\bibinfo {volume} {6}},\ \bibinfo {pages}
  {e1000783} (\bibinfo {year} {2010})}\BibitemShut {NoStop}%
\bibitem [{\citenamefont {Yang}\ \emph {et~al.}(2010)\citenamefont {Yang},
  \citenamefont {Rhind},\ and\ \citenamefont {Bechhoefer}}]{Yang2010}%
  \BibitemOpen
  \bibfield  {author} {\bibinfo {author} {\bibfnamefont {S.~C.-H.}\
  \bibnamefont {Yang}}, \bibinfo {author} {\bibfnamefont {N.}~\bibnamefont
  {Rhind}}, \ and\ \bibinfo {author} {\bibfnamefont {J.}~\bibnamefont
  {Bechhoefer}},\ }\href {\doibase 10.1038/msb.2010.61} {\bibfield  {journal}
  {\bibinfo  {journal} {Mol. Syst. Biol.}\ }\textbf {\bibinfo {volume} {6}},\
  \bibinfo {pages} {404} (\bibinfo {year} {2010})}\BibitemShut {NoStop}%
\bibitem [{\citenamefont {Nieduszynski}\ \emph {et~al.}(2007)\citenamefont
  {Nieduszynski} \emph {et~al.}}]{Nieduszynski2007}%
  \BibitemOpen
  \bibfield  {author} {\bibinfo {author} {\bibfnamefont {C.~A.}\ \bibnamefont
  {Nieduszynski}} \emph {et~al.},\ }\href {\doibase 10.1093/nar/gkl758}
  {\bibfield  {journal} {\bibinfo  {journal} {Nucl. Acids Res.}\ }\textbf
  {\bibinfo {volume} {35}},\ \bibinfo {pages} {D40} (\bibinfo {year}
  {2007})}\BibitemShut {NoStop}%
\bibitem [{\citenamefont {Mahbubani}\ \emph {et~al.}(1997)\citenamefont
  {Mahbubani} \emph {et~al.}}]{Mahbubani1997}%
  \BibitemOpen
  \bibfield  {author} {\bibinfo {author} {\bibfnamefont {H.~M.}\ \bibnamefont
  {Mahbubani}} \emph {et~al.},\ }\href {\doibase 10.1083/jcb.136.1.125}
  {\bibfield  {journal} {\bibinfo  {journal} {J. Cell. Biol.}\ }\textbf
  {\bibinfo {volume} {136}},\ \bibinfo {pages} {125} (\bibinfo {year}
  {1997})}\BibitemShut {NoStop}%
\bibitem [{\citenamefont {Blow}\ \emph {et~al.}(2001)\citenamefont {Blow} \emph
  {et~al.}}]{Blow2001a}%
  \BibitemOpen
  \bibfield  {author} {\bibinfo {author} {\bibfnamefont {J.~J.}\ \bibnamefont
  {Blow}} \emph {et~al.},\ }\href
  {http://www.pubmedcentral.nih.gov/articlerender.fcgi?artid=2193667\&tool=pmcentrez\&rendertype=abstract}
  {\bibfield  {journal} {\bibinfo  {journal} {J. Cell. Biol.}\ }\textbf
  {\bibinfo {volume} {152}},\ \bibinfo {pages} {15} (\bibinfo {year}
  {2001})}\BibitemShut {NoStop}%
\bibitem [{\citenamefont {Edwards}\ \emph {et~al.}(2002)\citenamefont {Edwards}
  \emph {et~al.}}]{Edwards2002}%
  \BibitemOpen
  \bibfield  {author} {\bibinfo {author} {\bibfnamefont {M.~C.}\ \bibnamefont
  {Edwards}} \emph {et~al.},\ }\href
  {http://www.ncbi.nlm.nih.gov/pubmed/12087101} {\bibfield  {journal} {\bibinfo
   {journal} {J. Biol. Chem.}\ }\textbf {\bibinfo {volume} {277}},\ \bibinfo
  {pages} {33049} (\bibinfo {year} {2002})}\BibitemShut {NoStop}%
\bibitem [{\citenamefont {Jun}\ and\ \citenamefont
  {Bechhoefer}(2005)}]{Jun2004a}%
  \BibitemOpen
  \bibfield  {author} {\bibinfo {author} {\bibfnamefont {S.}~\bibnamefont
  {Jun}}\ and\ \bibinfo {author} {\bibfnamefont {J.}~\bibnamefont
  {Bechhoefer}},\ }\href {\doibase 10.1103/PhysRevE.71.011909} {\bibfield
  {journal} {\bibinfo  {journal} {Physical Review E}\ }\textbf {\bibinfo
  {volume} {71}},\ \bibinfo {pages} {011909} (\bibinfo {year}
  {2005})}\BibitemShut {NoStop}%
\bibitem [{\citenamefont {Zhang}\ and\ \citenamefont
  {Bechhoefer}(2006)}]{Zhang2006}%
  \BibitemOpen
  \bibfield  {author} {\bibinfo {author} {\bibfnamefont {H.}~\bibnamefont
  {Zhang}}\ and\ \bibinfo {author} {\bibfnamefont {J.}~\bibnamefont
  {Bechhoefer}},\ }\href {http://www.ncbi.nlm.nih.gov/pubmed/16802963}
  {\bibfield  {journal} {\bibinfo  {journal} {Phys. Rev. E.}\ }\textbf
  {\bibinfo {volume} {73}},\ \bibinfo {pages} {051903} (\bibinfo {year}
  {2006})}\BibitemShut {NoStop}%
\bibitem [{\citenamefont {Goldar}\ \emph {et~al.}(2008)\citenamefont {Goldar}
  \emph {et~al.}}]{Goldar2008}%
  \BibitemOpen
  \bibfield  {author} {\bibinfo {author} {\bibfnamefont {A.}~\bibnamefont
  {Goldar}} \emph {et~al.},\ }\href {\doibase 10.1371/journal.pone.0002919}
  {\bibfield  {journal} {\bibinfo  {journal} {PLoS ONE}\ }\textbf {\bibinfo
  {volume} {3}},\ \bibinfo {pages} {e2919} (\bibinfo {year}
  {2008})}\BibitemShut {NoStop}%
\bibitem [{\citenamefont {Yang}\ and\ \citenamefont
  {Bechhoefer}(2008)}]{Yang2008}%
  \BibitemOpen
  \bibfield  {author} {\bibinfo {author} {\bibfnamefont {S.~C.-H.}\
  \bibnamefont {Yang}}\ and\ \bibinfo {author} {\bibfnamefont {J.}~\bibnamefont
  {Bechhoefer}},\ }\href {http://www.ncbi.nlm.nih.gov/pubmed/18999465}
  {\bibfield  {journal} {\bibinfo  {journal} {Phys. Rev. E.}\ }\textbf
  {\bibinfo {volume} {78}},\ \bibinfo {pages} {41917} (\bibinfo {year}
  {2008})}\BibitemShut {NoStop}%
\bibitem [{\citenamefont {Blow}\ and\ \citenamefont {Ge}(2009)}]{Blow2009}%
  \BibitemOpen
  \bibfield  {author} {\bibinfo {author} {\bibfnamefont {J.~J.}\ \bibnamefont
  {Blow}}\ and\ \bibinfo {author} {\bibfnamefont {X.~Q.}\ \bibnamefont {Ge}},\
  }\href {\doibase 10.1038/embor.2009.5} {\bibfield  {journal} {\bibinfo
  {journal} {EMBO reports}\ }\textbf {\bibinfo {volume} {10}},\ \bibinfo
  {pages} {406} (\bibinfo {year} {2009})}\BibitemShut {NoStop}%
\bibitem [{\citenamefont {Jun}\ \emph {et~al.}(2004)\citenamefont {Jun} \emph
  {et~al.}}]{Jun2004b}%
  \BibitemOpen
  \bibfield  {author} {\bibinfo {author} {\bibfnamefont {S.}~\bibnamefont
  {Jun}} \emph {et~al.},\ }\href {\doibase 10.4161/cc.3.2.655} {\bibfield
  {journal} {\bibinfo  {journal} {Cell Cycle}\ }\textbf {\bibinfo {volume}
  {3}},\ \bibinfo {pages} {211} (\bibinfo {year} {2004})}\BibitemShut {NoStop}%
\bibitem [{\citenamefont {PGAPack}()}]{PGAPack}%
  \BibitemOpen
  \bibfield  {author} {\bibinfo {author} {\bibnamefont {PGAPack}},\ }\href
  {http://ftp.mcs.anl.gov/pub/pgapack/} {\bibinfo  {journal}
  {http://ftp.mcs.anl.gov/pub/pgapack/}\ }\BibitemShut {NoStop}%
\bibitem [{\citenamefont {Shirahige}\ \emph {et~al.}(1993)\citenamefont
  {Shirahige} \emph {et~al.}}]{Shirahige1993}%
  \BibitemOpen
\bibfield  {journal} {  }\bibfield  {author} {\bibinfo {author} {\bibfnamefont
  {K.}~\bibnamefont {Shirahige}} \emph {et~al.},\ }\href
  {http://www.pubmedcentral.nih.gov/articlerender.fcgi?artid=360155\&tool=pmcentrez\&rendertype=abstract}
  {\bibfield  {journal} {\bibinfo  {journal} {Mol. Cell Biol.}\ }\textbf
  {\bibinfo {volume} {13}},\ \bibinfo {pages} {5043} (\bibinfo {year}
  {1993})}\BibitemShut {NoStop}%
\bibitem [{Note1()}]{Note1}%
  \BibitemOpen
  \bibinfo {note} {$F=\protect \frac {1}{9} \DOTSB \sum@ \slimits@ _
  {i=1}^{n=9}d_i^o/d_i^r$ is a measure of the difference between the gap
  distribution of the optimized and random cases. A gap is defined as the
  separation between the $i^\protect \text {th}$ experimental locus position
  $p_i^e$ and that of the optimization $p_i^o$: $d_i^o=|p_i^e-p_i^o|$. $d_i^r$
  is akin; the average separation that arises from placing a locus uniformly and 
  randomly and $p_i^e$. $F=0$ means that the optimization fits the experimental
  loci positions perfectly; $F\sim 1$ indicates no difference to that of a
  random placement.}\BibitemShut {Stop}%
\bibitem [{\citenamefont {Raghuraman}\ \emph {et~al.}(2001)\citenamefont
  {Raghuraman} \emph {et~al.}}]{Raghuraman2001}%
  \BibitemOpen
  \bibfield  {author} {\bibinfo {author} {\bibfnamefont {M.~K.}\ \bibnamefont
  {Raghuraman}} \emph {et~al.},\ }\href {\doibase 10.1126/science.294.5540.115}
  {\bibfield  {journal} {\bibinfo  {journal} {Science}\ }\textbf {\bibinfo
  {volume} {294}},\ \bibinfo {pages} {115} (\bibinfo {year}
  {2001})}\BibitemShut {NoStop}%
\bibitem [{\citenamefont {Herrick}(2002)}]{Herrick2002a}%
  \BibitemOpen
  \bibfield  {author} {\bibinfo {author} {\bibfnamefont {J.}~\bibnamefont
  {Herrick}},\ }\href {\doibase 10.1016/S0022-2836(02)00522-3} {\bibfield
  {journal} {\bibinfo  {journal} {J. Mol. Biol.}\ }\textbf {\bibinfo {volume}
  {320}},\ \bibinfo {pages} {741} (\bibinfo {year} {2002})}\BibitemShut
  {NoStop}%
\bibitem [{\citenamefont {Sekedat}\ \emph {et~al.}(2010)\citenamefont {Sekedat}
  \emph {et~al.}}]{Sekedat2010}%
  \BibitemOpen
  \bibfield  {author} {\bibinfo {author} {\bibfnamefont {M.~D.}\ \bibnamefont
  {Sekedat}} \emph {et~al.},\ }\href {\doibase 10.1038/msb.2010.8} {\bibfield
  {journal} {\bibinfo  {journal} {Mol. Syst. Biol.}\ }\textbf {\bibinfo
  {volume} {6}},\ \bibinfo {pages} {353} (\bibinfo {year} {2010})}\BibitemShut
  {NoStop}%
\bibitem [{\citenamefont {Rowles}\ \emph {et~al.}(1999)\citenamefont {Rowles},
  \citenamefont {Tada},\ and\ \citenamefont {Blow}}]{Rowles1999}%
  \BibitemOpen
  \bibfield  {author} {\bibinfo {author} {\bibfnamefont {A.}~\bibnamefont
  {Rowles}}, \bibinfo {author} {\bibfnamefont {S.}~\bibnamefont {Tada}}, \ and\
  \bibinfo {author} {\bibfnamefont {J.~J.}\ \bibnamefont {Blow}},\ }\href
  {http://www.ncbi.nlm.nih.gov/pubmed/10341218} {\bibfield  {journal} {\bibinfo
   {journal} {Journal of cell science}\ }\textbf {\bibinfo {volume} {112}},\
  \bibinfo {pages} {2011} (\bibinfo {year} {1999})}\BibitemShut {NoStop}%
\bibitem [{\citenamefont {Oehlmann}\ \emph {et~al.}(2004)\citenamefont
  {Oehlmann}, \citenamefont {Score},\ and\ \citenamefont
  {Blow}}]{Oehlmann2004}%
  \BibitemOpen
  \bibfield  {author} {\bibinfo {author} {\bibfnamefont {M.}~\bibnamefont
  {Oehlmann}}, \bibinfo {author} {\bibfnamefont {A.~J.}\ \bibnamefont {Score}},
  \ and\ \bibinfo {author} {\bibfnamefont {J.~J.}\ \bibnamefont {Blow}},\
  }\href {\doibase 10.1083/jcb.200311044} {\bibfield  {journal} {\bibinfo
  {journal} {The Journal of cell biology}\ }\textbf {\bibinfo {volume} {165}},\
  \bibinfo {pages} {181} (\bibinfo {year} {2004})}\BibitemShut {NoStop}%
\end{thebibliography}
\end{document}